\documentclass[
    ,final            
  ]
  {aipproc}
  
\usepackage{graphicx}
\layoutstyle{6x9}

\def\ra{\rightarrow}

\def\be{\begin{equation}}
\def\ee{\end{equation}}
\def\bea{\begin{eqnarray}}
\def\eea{\end{eqnarray}}

\begin{document}

\title{$g_{1}(x)$ and $g_{2}(x)$ in the Meson Cloud Model}


\author{A. I. Signal}{
  address={Institute of Fundamental Sciences PN461 \\ Massey University \\
Private Bag 11 222,  Palmerston North \\
New Zealand}
}



\begin{abstract}
 We calculate the spin dependent structure functions $g_{1}(x)$ and $g_{2}(x)$ 
of the proton and neutron.
Our calculation uses the meson cloud model of nucleon structure and includes 
the effects of kinematic terms which mix transverse and longitudinal spin components.
We find small corrections to the nucleon structure functions, however these are 
significant for the neutron.
\end{abstract}

\maketitle


The spin dependent structure functions of the nucleon are the subject of much 
theoretical and experimental interest. 
As deep inelastic (and other) experiments become more precise it is hoped that it 
may be possible to make an unequivocal measurement of a higher twist component 
in the structure function $g_{2}$ of the nucleon. 
This would give new information on the gluon field inside the nucleon, and its 
relationship with the quark fields.  

In order to make such an unequivocal identification it is necessary to understand 
the relationship between the structure functions $g_{1}$ and $g_{2}$. 
In particular there are leading twist contributions to $g_{2}$ which arise from 
scattering from the components of the meson cloud of the physical nucleon.
These contributions are of the order of 10\% of the structure function, and need 
to be taken into account when calculating the twist-2 part of $g_{2}$.

The Meson Cloud Model (MCM) arises from the crucial observation \cite{Sullivan} 
that the contribution of scattering from the pion cloud of the nucleon scales in the 
Bjorken limit. 
This implies that the parton distributions of the nucleon are modified via a convolution 
between the parton distribution of the meson and the momentum distribution of the meson 
in the proton, viz.
\bea
\delta q^{p}(x) = \int_{x}^{1} \frac{dy}{y} f_{p \pi}(y) q^{\pi}\left(\frac{x}{y}\right).
\eea
As well as pions, the MCM takes into account scattering from the other baryon plus meson 
components in the Fock expansion of the wavefunction i.e.
\bea
|N\rangle_{\rm physical} =  \sqrt{Z} |N\rangle_{\rm bare}
+\sum_{MB}  
\int dy \, d^2 {\bf k}_\perp \, \phi(y,k_\perp^2)
\, |M(y, {\bf k}_\perp); B(1-y,-{\bf k}_\perp)
\rangle. 
\eea
The other ingredients of the model are the interaction Lagrangians ${\cal L}_{int}$ 
describing the $N \ra BM$ vertices and the form factors for these vertices.
The small probability of finding high mass states in this model leads to quick 
convergence of the sum over baryon-meson states for structure function calculations.

The MCM has been applied successfully in spin independent DIS, giving a good description 
of the HERA data on semi-inclusive DIS with a leading neutron \cite{H1n, Zeusn}, and 
also dijet events with a leading neutron \cite{Zeus2j, H12j}. 
In addition the MCM gives a good description of the observed violation of the Gottfried 
sum rule \cite{NMC, E866}.

To extend the model to spin dependent DIS requires the contributions of both pseudoscalar 
and pseudovector mesons, particularly the $\rho$ meson.
The pseudoscalar contributions mainly `dilute' the bare spin dependent pdfs, however the 
importance of $L \neq 0$ amplitudes in the cloud cannot be ignored \cite{HHoltmannSS, BorosT}. 
The pseudovector mesons can contribute directly to the spin dependent pdfs. 
In earlier work we calculated the spin dependent sea distributions 
$\Delta \bar{u}(x), \;  \Delta \bar{d}(x), \; s(x)$ and $\Delta \bar{s}(x)$ \cite{FCaoS_ps2}. 
Our results are in good agreement with the HERMES data \cite{HERMES02}.

The structure functions $g_{1}(x)$ and $g_{2}(x)$ are dominated by valence rather than sea 
distributions, so the most important contributions in the MCM are those affecting the valence 
quarks, which are $N \ra N \pi$ and $N \ra \Delta \pi$, with 
${\cal L}_{int} = i g_{NN\pi} \bar{\psi} \gamma_{5} \pi \psi, \; 
f_{N \Delta \pi} \bar{\psi} \partial_{\mu} \pi\chi^{\mu} + \mbox{h.c.}$ respectively. 

At finite $Q^{2}$ the spin of the struck hadron from the cloud, in this case either a 
nucleon or $\Delta$, is not parallel with the initial spin of the target nucleon. 
This implies,as shown by Kumano and Miyama \cite{KM}, that both longitudinal and transverse 
spin structure functions of the cloud hadrons contribute to the observed structure functions.
For a spin $1/2$ baryon component of the cloud we have
\bea
\delta g_{1}(x, Q^{2}) & = & \frac{1}{1+\gamma^{2}} \int_{x}^{1} \frac{dy}{y} \sum_{i = 1,2} 
(-1)^{i+1}[\Delta f_{i L}(y) + \Delta f_{i T}(y)] g_{i}^{B}(\frac{x}{y}, Q^{2}) \\
\delta g_{2}(x, Q^{2}) & = & \frac{1}{1+\gamma^{2}} \int_{x}^{1} \frac{dy}{y} \sum_{i = 1,2} 
(-1)^{i} \left[\Delta f_{i L}(y) + \frac{\Delta f_{i T}(y)}{\gamma^{2}} \right] g_{i}^{B}(\frac{x}{y}, Q^{2}).
\label{eq:dgs}
\eea
where $\gamma^{2} = 4x^{2}m_{N}^{2} / Q^{2}$ and $\Delta f_{i L,T}(y)$ are the diferences 
between spin up and spin down fluctuation functions projected longitudinally along or 
transverse to the baryon 3-momentum.
Similar expressions exist for higher spin components of the cloud \cite{KM, BCS}.
The fluctuations are calculated using standard techniques in time-ordered perturbation 
theory in the infinite momentum frame \cite{HHoltmannSS, KM, BCS}. 
We find that for longitudinal fluctuation functions $\Delta f_{i L}(y)$ the nucleon and 
$\Delta$ contributions are of similar size, with the $s = 3/2$ state of the $\Delta$ being 
important.
For transverse fluctuations $\Delta f_{i T}(y)$ the nucleon contributions are much larger 
than those from the $\Delta$.

In order to estimate the size of the MCM contributions to $g_{1}$ and $g_{2}$, we need also 
to calculate the structure functions of the `bare' hadrons. 
We use the MIT bag model and the methods developed by the Adelaide group 
\cite{BorosT, Bag_Adelaide} to calculate the spin dependent pdfs. 
We also add `by hand' a phenomenological $\Delta g(x)$ such that the integral of 
$g_{1}^{p}(x)$ agrees with experiment.
The resulting $g_{1}^{p}(x)$ and $g_{1}^{n}(x)$ give a reasonable description of the 
experimental data. 
To calculate the bare $g_{2}(x)$ we simply use the leading twist Wandzura-Wilczek 
term
\bea
g_{2}(x) = -g_{1}(x) + \int_{x}^{1} \frac{dy}{y} g_{1}(y),
\eea
which also gives a good description of the available experimental data on $g_{2}^{p}(x)$ 
and $g_{2}^{n}(x)$.

Our calculations of the contributions to $g_{1}(x)$ and $g_{2}(x)$ for both the proton 
and neutron are shown in figures 1 and 2. 
For the proton we note that the magnitudes of these contributions are much smaller than the 
size of the experimental data. 
We see that the contributions from transversely projected cloud baryons are small, and 
that the contributions of nucleons and $\Delta$ baryons are of similar magnitude, though not 
necessarily the same sign. 
For the neutron structure functions the MCM contributions are around 10\% of the 
size of the experimental data. 
These corrections will be important to consider in any extraction of higher twist components 
to the neutron structure functions, as they have similar magnitude to these components. 
Also these corrections have a weak scale dependence which can mimic that of  twist-3
contributions at low $Q^{2}$.
The correction to $g_{2}^{n}(x)$ is positive, and may  be able  to account for the deviation 
of the JLab E97-103 data  around $x = 0.2$ and $Q^{2} = 1$ GeV$^{2}$ from the 
Wandzura-Wilczek term \cite{Cheng}.

\begin{figure}
\centering
  \includegraphics[width=2.5in]{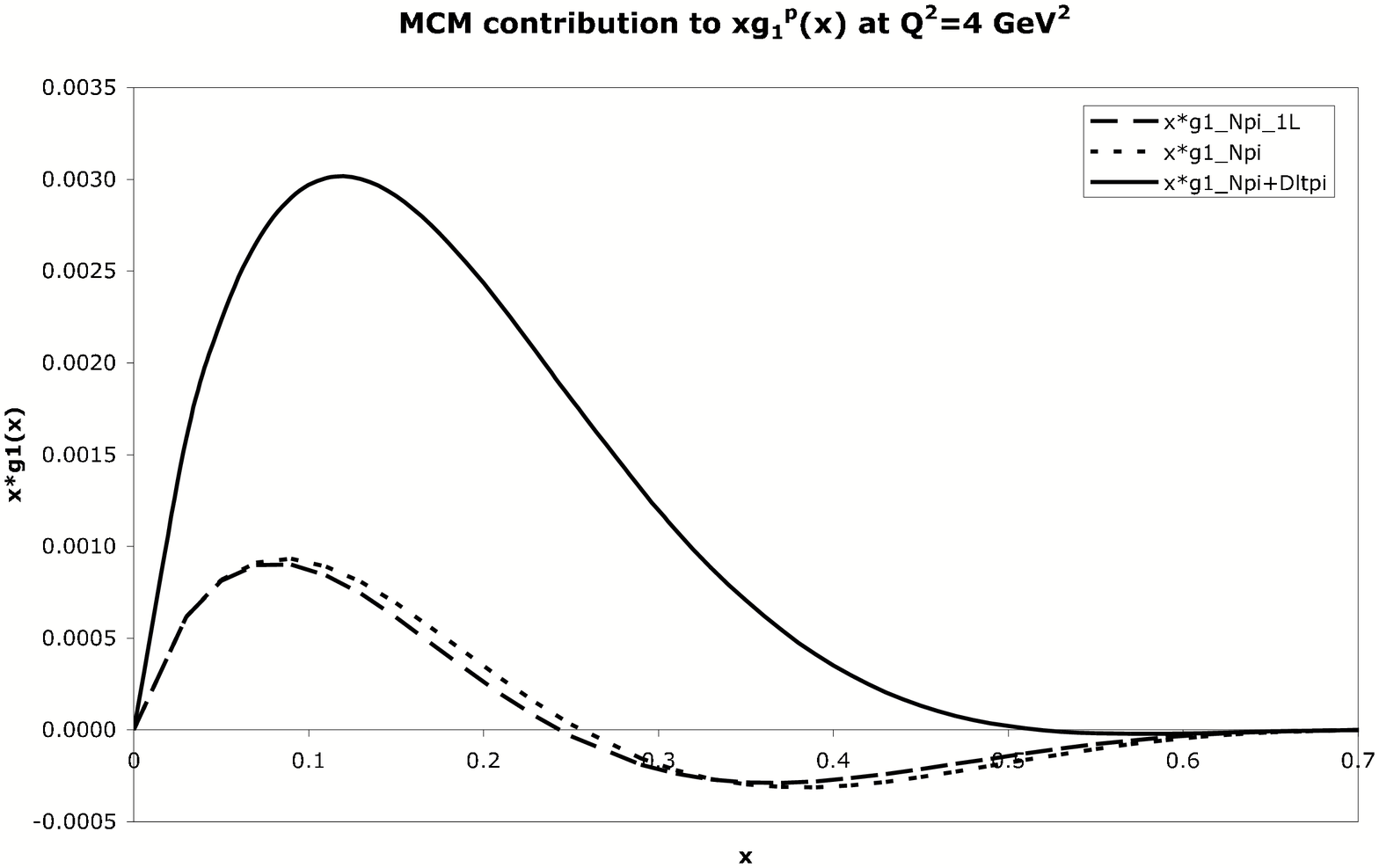}
  \hfill
  \includegraphics[width=2.5in]{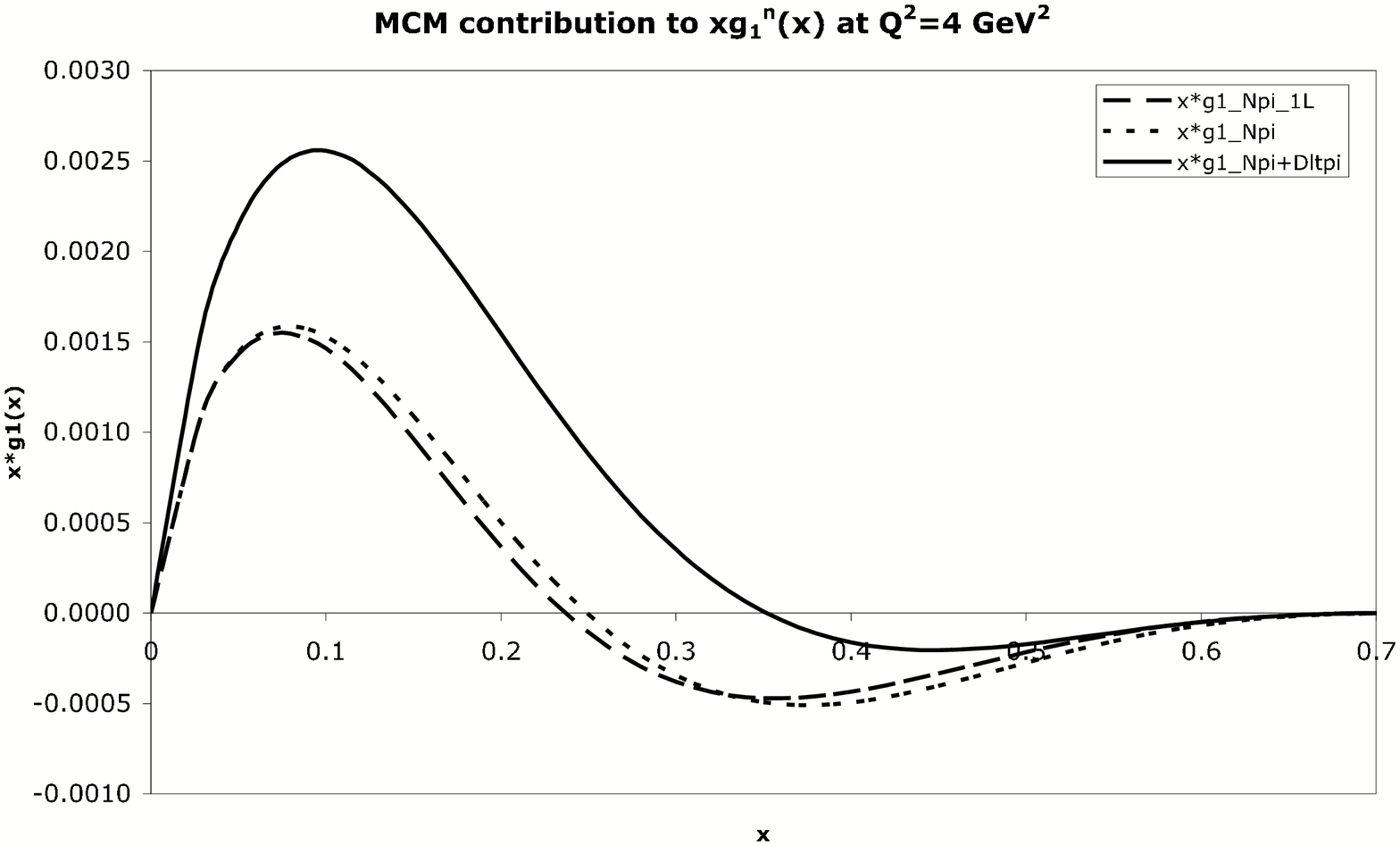}  
  \caption{Meson cloud Model contributions to $g_{1}$ of the proton and neutron. 
  	The dashed line is the contribution from longitudinally projected nucleon fluctuations, 
	the dotted line is the total contribution from nucleon fluctuations and the solid line is 
	the total from nucleon and $\Delta$ fluctuations.}
\end{figure}

\begin{figure}
\centering
  \includegraphics[width=2.5in]{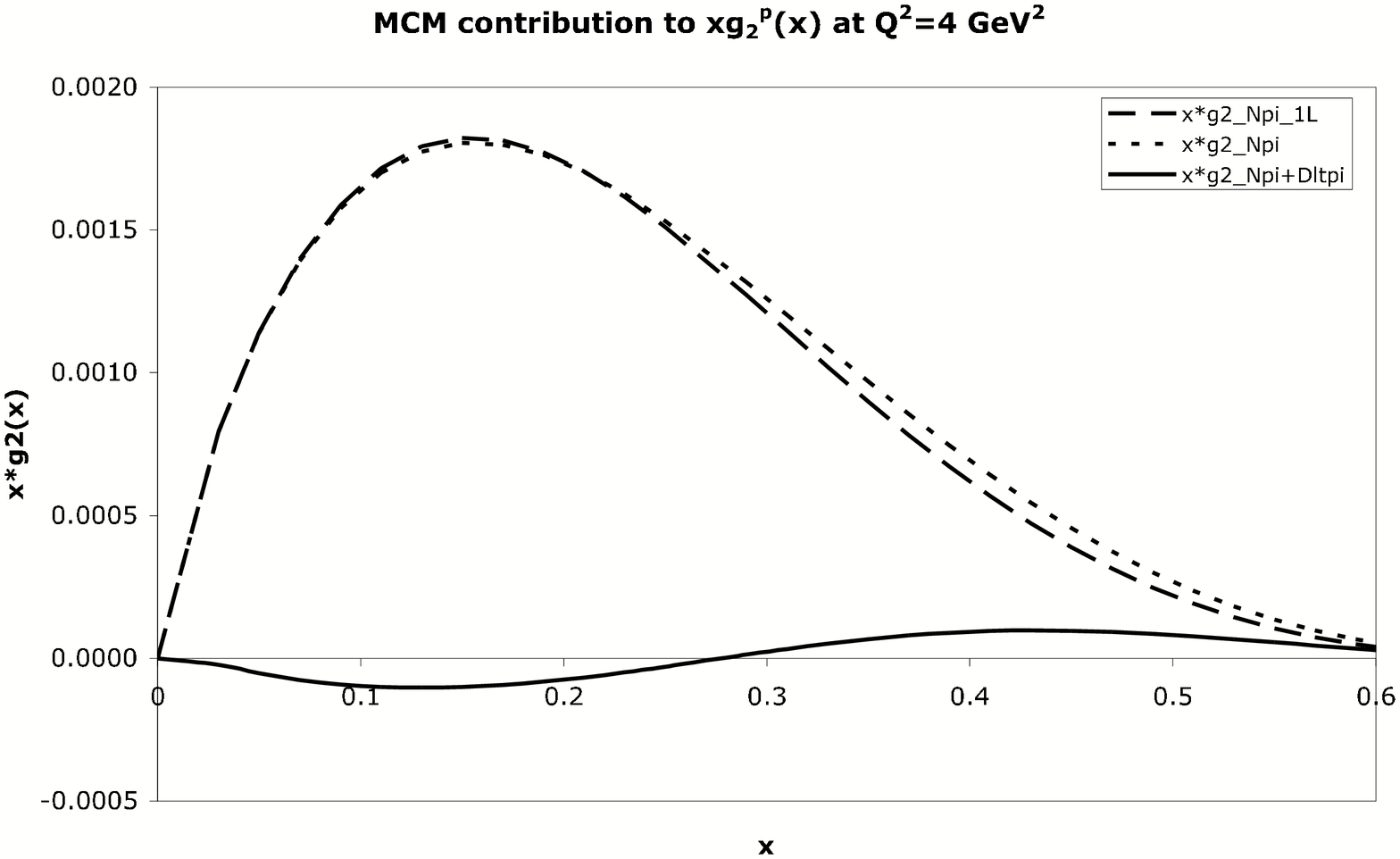}
    \hfill
  \includegraphics[width=2.5in]{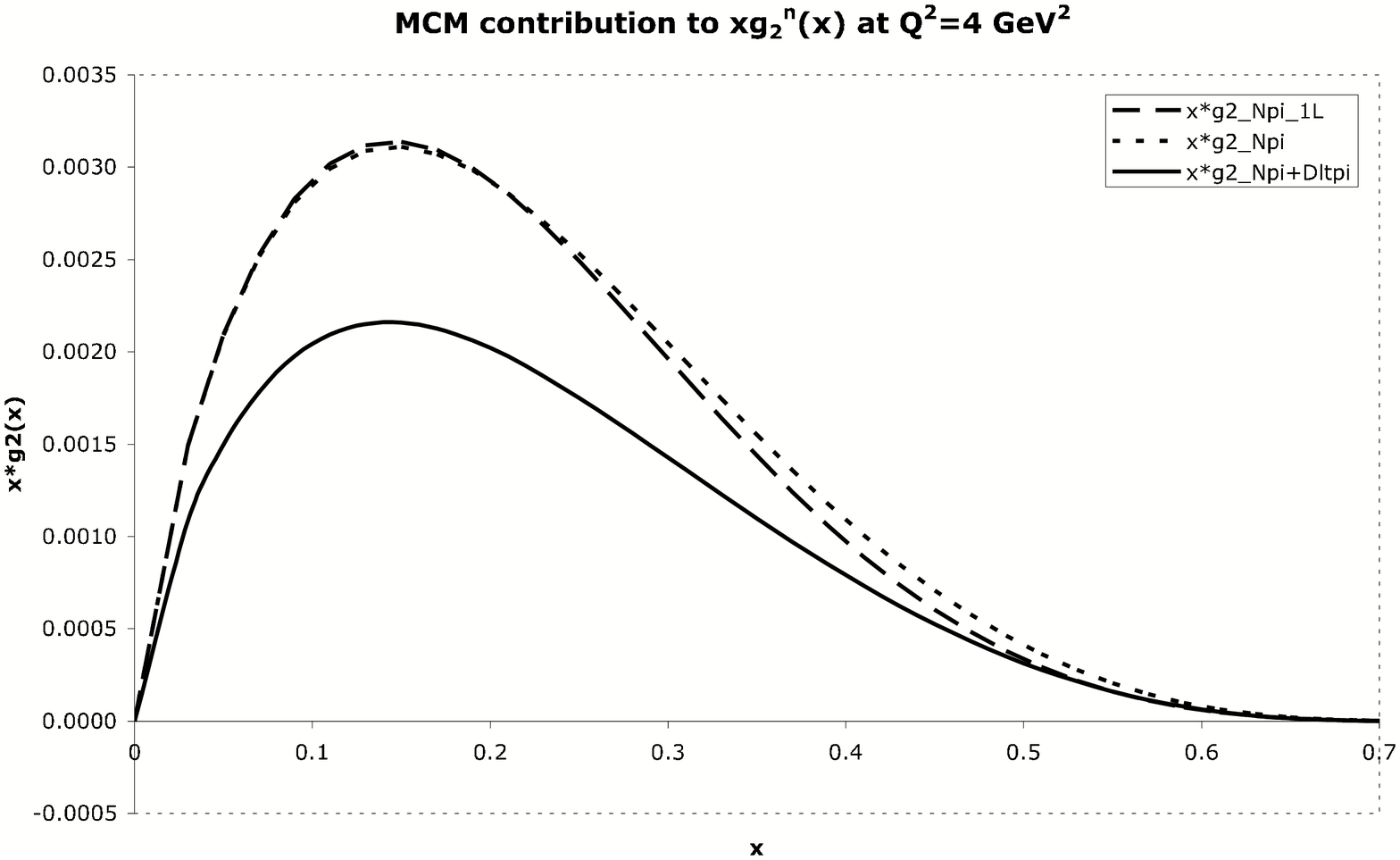}  
  \caption{Meson cloud Model contributions to $g_{2}$ of the proton and neutron. 
  	The dashed line is the contribution from longitudinally projected nucleon fluctuations, 
	the dotted line is the total contribution from nucleon fluctuations and the solid line is 
	the total from nucleon and $\Delta$ fluctuations.}
\end{figure}



\begin{theacknowledgments}
A.S. acknowledges the hospitality and support of the Institute for Particle Physics
Phenomenology, Durham University, where portions of this work were done.
\end{theacknowledgments}



\bibliographystyle{aipproc}   


\begin{thebibliography}{9}

\bibitem{Sullivan}
	J. D. Sullivan, \emph{Phys. Rev. D}, \textbf{5}, 1732 (1972).
\bibitem{H1n}
	C. Adloff et al.,  \emph{Eur. Phys. J. C}, \textbf{6}, 587 (1999).
\bibitem{Zeusn}
	M. Derrick et al.,  \emph{Phys. Lett. B}, \textbf{384}, 388 (1996);
	S. Chekanov et al., \emph{Nucl. Phys. B}, \textbf{637}, 3 (2002).
\bibitem{Zeus2j}
	J. Breitweg et al.,  \emph{Nucl. Phys. B}, \textbf{596}, 3 (2001).
\bibitem{H12j}
	A. Aktas et al.,  hep-ex/0501074, \emph{submitted to Eur. Phys. J. C.}
\bibitem{NMC}
	P. Amaudraz et al., \emph{Phys. Rev. Lett.}, \textbf{66}, 2712 (1991);
	M. Arneodo et al., \emph{Phys. Rev. D}, \textbf{50}, 1 (1994);
	M. Arneodo et al.,  \emph{Phys. Lett. B}, \textbf{364}, 107 (1995).
\bibitem{E866}
	E. A. Hawker et al., \emph{Phys. Rev. Lett.}, \textbf{80}, 3715 (1998);
	\emph{Phys. Rev. D}, \textbf{64}, 052002 (2001).
\bibitem{HHoltmannSS}
	H. Holtmann, A. Szczurek, and J. Speth, \emph{Nucl. Phys. A}, \textbf{569}, 631 (1996).
\bibitem{BorosT}
	C. Boros and A. W. Thomas, \emph{Phys. Rev. D}, \textbf{60}, 074017 (1999).
\bibitem{FCaoS_ps2}
	F. G. Cao and A. I. Signal, \emph{Phys. Rev. D}, \textbf{68}, 074002 (2003).
\bibitem{HERMES02}
	A. Airapetian et al., \emph{Phys. Rev. Lett.}, \textbf{92}, 012005 (2004). 
\bibitem{KM}
	S. Kumano and M. Miyama, \emph{Phys. Rev. D}, \textbf{65}, 034012 (2002).
\bibitem{BCS}
	F. Bissey, F. G. Cao and A. I. Signal, \emph{in preparation}.
\bibitem{Bag_Adelaide}
	A. I. Signal and A. W. Thomas, \emph{Phys. Lett. B}, \textbf{211}, 481 (1988);
	\emph{Phys. Rev. D}, \textbf{40}, 2832 (1989);
	A. W. Schreiber, A. W. Thomas, and J. T. Londergan, \emph{Phys. Rev. D}, \textbf{42}, 2226 (1990).
\bibitem{Cheng}
	J. P. Cheng, \emph{presentation at DIS05}.	
\end{thebibliography}





\end{document}